\documentclass[aip,amsmath,amssymb,reprint,aps,prb,citeautoscript,nofootinbib]{revtex4}%{achemso}
\usepackage[dvips]{graphicx}
\usepackage{multirow}
\usepackage{array}
\usepackage{etoolbox}
\usepackage{inputenc}
\begin{document}

\title{Disordered peptide chains in an $\alpha$-C-based  coarse-grained model}

\author{{\L}ukasz Mioduszewski}
\affiliation{Institute of Physics, Polish Academy of Sciences,
 Al. Lotnik{\'o}w 32/46, 02-668 Warsaw, \\ Poland}

\author{Marek Cieplak\footnote{E-mail: mc@ifpan.edu.pl}}
%\email{mc@ifpan.edu.pl}
\affiliation{Institute of Physics, Polish Academy of Sciences,
 Al. Lotnik{\'o}w 32/46, 02-668 Warsaw, \\ Poland}
\date{\today}

\begin{abstract}
\noindent
We construct a one-bead-per-residue coarse-grained dynamical model
to describe intrinsically disordered proteins at significantly longer
timescales than in the all-atom models. In this model,
the inter-residue contacts form and disappear during the course
of the time evolution. The contacts may arise between the sidechains,
the backbones and the sidechains and backbones of the interacting residues.
The model yields
results that are consistent with many all-atom and experimental
data on these systems. We demonstrate that the geometrical properties
of various homopeptides differ substantially in this model. In particular,
the average radius of gyration scales with the sequence length in a
residue-dependent manner.
\end{abstract}

\maketitle

\section{Introduction}

Coarse-grained models 
\cite{Warshel,Levitt,Kolinski,Liwo,Tozzini,Kim} have provided valuable computational
tools to gain insights into the temperature- or force-induced dynamics
of large conformational transformations of proteins. These models have been applied
mostly to proteins that are natively structured, usually globular, and they
often consider the solvent to be implicit. One convenient way to
implement coarse-graining is to adopt the structure-based approach
\cite{Go,Vishv,Takada,Clementi,Hoang1} in which the dynamics are governed
by contact interactions and the parameters involved
are derived so that the native state coincides with the
ground state of the system. This idea embodies the principle of
minimal frustration and leads to an optimal
folding funnel \cite{Bryngelson,Baker}.\\

The issues of conformational dynamics become center-stage when describing 
intrinsically disordered proteins (IDPs) \cite{Dyson,Fink,Uversky,Ferreon,Uversky1}.
This is because such systems evolve through various competing basins of
attraction instead of staying mostly within one deep valley
in the free energy landscape, corresponding to the native state. 
Some IDP systems, for instance
$\alpha$-synuclein \cite{Gnanakaran}, polyvaline (polyV) \cite{Cossio} and
polyglutamine (polyQ) \cite{Vitalis,Esposito,Ogawa,Gomez}, have been studied by
all-atom simulations, despite persistent questions about the validity of the
standard force fields in this case \cite{Rauscher,Chen,Nawrocki}.
However, the all-atom studies come with limited time-scales
of the simulations and thus with restricted probing of the possible basins
(this problem can be alleviated somewhat by eliminating the explicit solvent,
by using GPU clusters and by shortening folding trajectories 
for structured proteins by involving effective polarizable force
fields \cite{Duan}.
In addition, it is likely that on very long time scales many of the
all-atom details of description cease to be relevant.
Thus there is a need to develop coarse-grained models for
molecular dynamics of the IDPs. %molecular dynamics approaches for the IDPs. 
\\

The existing approaches are based on the Monte Carlo method \cite{Kim},
Brownian dynamics \cite{coffdrop}, discontinuous molecular dynamics \cite{prime,Hall}
and molecular dynamics with an explicit solvent \cite{martini}.
All of them use more than one pseudo-atom per amino acid. In addition,
with the exception of the last approach, they cannot be
easily applied in situations in which
a protein is partially structured and partially disordered.  %, or, in the case of 
Also, the
existing hybrid Go-models with nonnative contacts \cite{DeSancho,Ganguly,Rey2} 
do not take into account directionality and distance dependence of the contacts.\\

Here, we construct a novel one-bead-per-residue top-down approach formulated
in the spirit of the structure-based models -- specifically
they involve contact interactions. The difference, however,
is that the parameters used are not determined for one specific 
protein, but by a statistical method, based on 
a non-redundant set of structures from Protein Data Bank 
as collected in the CATH database \cite{CATH}. 
The parameters, especially the values of characteristic lengths in
the contact potentials, incorporate amino acid specificity.
Another difference is that the establishment of contacts
between the $\alpha$-C atoms is governed not only by
their mutual distance but also by the expected directionality of the side
chains as derived from the positions of the $\alpha$-C atoms. 
In this respect, we borrow from several Monte Carlo studies pertaining
to the natively structured proteins \cite{tube,Buchete,Rey} and fibril 
formation \cite{Hoang} but provide a molecular dynamics implementation
of this concept. It should be pointed out that, unlike most other coarse-grained
approaches, the nature and characteristics of a given contact may be
time-dependent because an interaction between, say, two sidechains
may switch to that between one sidechain and the other backbone.
\\

We show that, despite the rather approximate nature of our model,
the approach yields results which are consistent with many
previous computational and experimental results
pertaining to the average geometry of the polypeptide chains
of various lengths, $n$. We also predict values of the geometrical
parameters for 20 homopeptides of the same chain length ($n=30$)
and demonstrate existence of substantial differences
between the homopeptides made of different residues.
We have found that in the UniProt Knowledge Database \cite{UniProt},
the naturally occurring homopolymeric tracts range in the maximal sequential 
length, $n_{max}$, between 6 (Ile, Tyr, Trp) and 79 (Gln). The full list of these maximal
lengths is given as one of the entries in Table \ref{aaproperties}.
Nevertheless, arbitrary lengths can be considered theoretically which
allows  us to determine the scaling exponents for the variations of the
average radii of gyration %, $R_g$, 
with $n$ and show that they depend on
the nature of the residue. It is expected that a chemical synthesis
can lead to the tract lengths that exceed $n_{max}$.\\

In this work, we consider only the case of monomers. However, the true 
strength of this coarse-grained approach is expected to show in studies
of many-chain disordered systems such as gluten \cite{gluten,gluten1}
and of large protein aggregates and complexes with disordered linkers
such as arising in cellulosomes \cite{Bayer,Miriam,linkers}.
These large systems are inaccessible to meaningful all-atom modelling
and assessing, say, the viscoelastic properties of gluten requires
very long simulation times.

\section{Methodology}

Our coarse-grained approach is, in many ways, %as outlined in refs. 
as outlined in our structure-based model \cite{JPCM,plos,models}
except that now there is no single native structure that could be
used to derive parameters in the potentials.
The peptide chains are represented by harmonically connected beads
that also participate in transient contact interactions. The beads are
located at the $\alpha$-C atoms that are at the distance, $r_b=3.8$ 
{\AA} apart. The contact interactions between
beads $i$ and $j$, separated by the distance of $r_{i,j}$,
are described by the Lennard-Jones (LJ) potential
$V_{L-J}(r_{i,j})=4\epsilon\left[{\left(\frac{\sigma_{i,j}}{r_{i,j}}\right)}^{12}-
\left(\frac{\sigma_{i,j}}{r_{i,j}}\right)^6 \right]$, 
where the energy parameter $\epsilon$ is assumed to be of the same order
as for the structured proteins: about
110 pN~{\AA} obtained by matching to the experimental data
on stretching \cite{plos}. This value is close to 1.5 kcal/mol obtained independently by
matching all-atom energies to the coarse-grained expressions \cite{Poma}.
We determine the length parameter $\sigma_{i,j}=r_{min}\cdot(0.5)^\frac{1}{6}$, 
(where $r_{min}$ is the distance where the potential acquires its
minimal value) based on the chemical identity of the
residues involved. Its choice and the criteria for the
contact formation will be discussed below.
The elastic constant in the tethering harmonic potential, mimicking the
peptide bonds, is set to $k=100$ \AA$^{-2}\cdot\epsilon$.\\

\subsection{Contact interactions}

In structure-based coarse-grained models, the contact map is
fixed and is determined by the native structure. All of these
contacts are occupied only in the native state, otherwise
the occupation is only partial.
In the case of the IDPs, there is no a priori fixed contact map.
However, at each instant, there is a set of contacts that
are active, i.e. they provide attraction between residues.\\

In order to learn how to determine which contacts are 
functional, we first make a survey of
21 090 structured proteins from the database provided by the CATH database \cite{CATH}
(the set of proteins with the sequence similarity not exceeding 40\%:
cath-dataset-nonredundant-S40.pdb) and
determine the nature of the native contacts as defined by the
overlap (OV) criterion that is discussed in more details in ref.
\cite{Wolek} Briefly, the heavy atoms are represented by spheres.
The radii of the spheres are equal to the context-dependent values as
listed in ref. \cite{Tsai} and multiplied by 1.24 to account for attraction
(the factor corresponds to the inflection point in the LJ 
potential) \cite{Settanni}.
If at least one sphere belonging to amino acid $i$ overlaps with
at least one sphere belonging to amino acid $j$ then we declare
existence of the native contact between $i$ and $j$. 
Unlike a procedure in which a contact is declared to be present
based on a cutoff distance, our criterion directly relates to the
conformation of the side groups and to the sizes of the atoms.\\

\begin{figure}[h]
 \centering
 \includegraphics[width=0.5\textwidth]{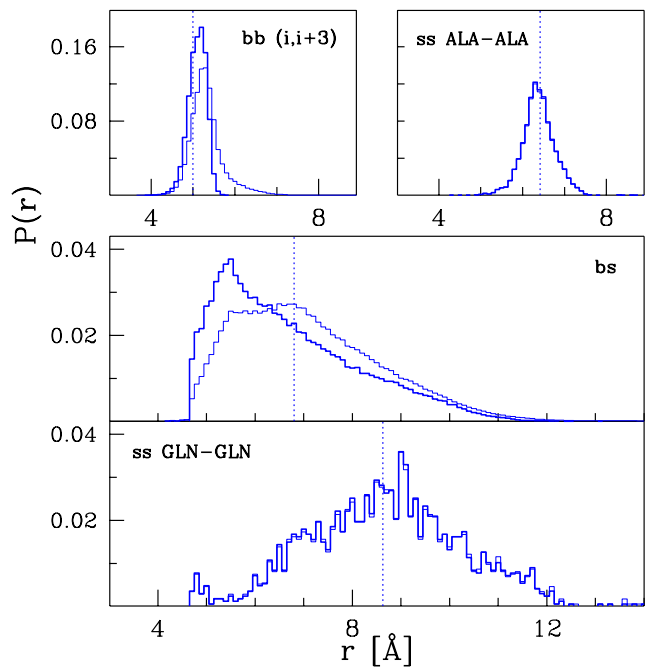}
 \caption{Examples of the distributions of the $\alpha$-C -- $\alpha$-C
distances in contacts. The distributions are for situations in which
not more than one contact is made between the atoms. The thin lines are
for all of these situations whereas the thick lines correspond to eliminating
contacts which do not satisfy the required directional conditions.
The vertical dotted lines indicate the values adopted in our definition
of the model. They correspond to the average value calculated based on the
thick histograms. The top left panel is for the bb contacts
regardless the specificity. Most of them
occur in the $i,i+3$ contacts so only those were taken into account. 
The average distance is about 5.0 {\AA}. However, the average distances
in other contacts are close in value: 4.7 {\AA} for the $i,i+4$ contacts
and 4.8 {\AA} for contacts at all larger sequential distances.
The middle panel is for the bs contacts, again regardless the specificity.
Almost all of such contacts occur in contacts $i,i+k$, where $k \ge 5$
and only these were included in the distribution.
The remaining panels are for the ss contacts for Ala-Ala (the top right panel)
and Gln-Gln (the bottom panel). These are determined for all sequential 
distances combined. Similar panels for all possible distributions 
corresponding to the
ss contacts are shown in the Electronic Supplementary Information (Figs. S1-S3).
 } 
 \label{distri}
 \end{figure}

We distinguish three classes of the contacts depending on
the residue fragment from which the OV-making spheres
originate: side chain - side chain (ss), 
backbone - backbone (bb) and backbone - side chain (bs). For a given
pair of amino acids there could be simultaneous overlaps
belonging to different classes. Independently of their number,
only one effective contact is used in the dynamical description.
It corresponds to a given $\alpha$-C -- $\alpha$-C distance.
We determine the distribution of these distances in the three
classes: bb, bs and ss  as illustrated in Fig. \ref{distri}.
However, in these distributions,
the presence of, say, an ss contact does not preclude
a simultaneous OV of the bs or bb kind in a given pair of residues.
In our model, though, for a given distance, only one kind of
contact can be operational for a pair of residues.
It should be noted that our effective description involves only the
$\alpha$-C atoms instead of introducing (up to) two kinds
of beads to represent a residue, one for the backbone atom and
another for a side chain, as proposed, for instance, by Gu {\it et al}. \cite{Gu}
%in ref. \cite{Gu}.\\

In order to associate a characteristic
distance with a unique type of contact, we derive subdistributions
corresponding to situations in which the OVs arise
only in one class (e.g., only ss).
In the case of the ss contacts, we also divide them into pairs
of specific residues as listed in Table \ref{rmin}. In our model, 
like-charged residues do not form any ss contacts. Neither do
glycine (Gly) and proline (Pro) as they can form only the backbone-involving
contacts. Thus, the table has 165 entries (instead of 210).\\

The numbers listed in the Table \ref{rmin} correspond to the values
of $r_{min}$ used. They range between 6.42 {\AA} (Ala--Ala) 
and 10.85 {\AA} (Trp--Trp).
In principle, they are equal to the
average values in the distribution. However, they are calculated by
also taking into account additional restrictions that relate to the
directionality, as discussed later on, though the corrections
due to these restrictions are minor.
The large value of $r_{min}$ for the Trp-Trp pairs requires
a special consideration
as it would lead to unphysical extended conformations in polyW tracts
that are not found in the PDB \cite{PDB}. The longest polyW tract with a known
structure is found in PDB:3N85. It has the length of 4 and it
corresponds to a turn. The extended conformations do not arise if the
$i,i+3$ ss contacts in the Trp-Trp pairs are disabled.\\

For the bb and bs
cases, we take $r_{min}$ of 5.0 and 6.8 {\AA}, respectively, without
incorporating any specificity.\\

%Following refs. \cite{Rey2,Hoang}
Following previous works \cite{Rey2,Hoang}, the bb contacts are treated in a 
special way: their well energy is set to $2\;\epsilon$ (energy of every other 
contact is $\epsilon$) and they cannot be formed between %the $i$th and $i+4$th residues. 
the $i,i+4$ pairs of residues.
This approach takes into account ,,renumbering'' \cite{renumbering} of the
backbone hydrogen bonds in a coarse-grained representation of an $\alpha$-helix: 
the $i,i+4$ bb contacts that have been identified by the OV criterion,
applied to globular structures, have the $\alpha$-C - $\alpha$-C distances
that are  close to 6 {\AA} (data not shown) and over 98\% of them are 
accompanied by a simultaneously arising bs contact. 
Therefore, it is sufficient to attribute the 
effective $i,i+4$ interaction to the bs-based contact.
In all other bb contacts, the typical distance is about 5 {\AA}
and an association with a bs contact is much less common.
Doubling the strength of the bb contacts compensates for the strict 
criteria for forming them, agrees with the earlier literature 
findings \cite{Rey2}  and conserves the balance between
the number of each type of contacts by making the bb contacts harder 
to break by thermal fluctuations (in structured proteins 
there are about twice as many ss contacts as the bb contacts \cite{Buchete}).
\\

The time evolution is based on the equations of motion
for the $\alpha$-C atoms. During its course,
an effective contact may appear or disappear based on the
distance between the $\alpha$-C atoms (as illustrated in Fig. \ref{arrows})
and on the shape of the conformation.
Switching the contacts on and off is done quasi-adiabatically. \\

At any instant, a contact between a given pair of residues
is allowed to arise only as a result of one mechanism,
corresponding to its specific value of $r_{min}$.
A new type of mechanism is allowed to act only after the
potential associated with the previous one drops to zero. Thus
before a new type of contact can be turned on, the previous type is 
completely turned off. In practice, dynamical switching between various
contact mechanisms occurs infrequently.
The electrostatic interactions 
are not considered to be contacts
in the sense used in our molecular dynamics model. 
It should be noted that if one uses the OV criterion,
then the various kinds of the overlap may act together
but -- in the structure-based model -- this results just
in single standard values of the energy and length parameters
of the contact, independent of the number of the
contributing OVs. Thus, in our dynamical model
two mechanisms cannot act together so that the contact does not
get described by a sum of, say, two potentials. 
However, it might be interesting to consider a model in which
the effective $r_{min}$  is a weighted average of the
length parameters corresponding to different successive mechanisms.
\\

All beads are endowed with the repulsive
potential to ensure that the chain never passes through itself. 
This potential is always turned on, regardless of the
possible additional presence of the attractive potentials.
The repulsion is described by the 
LJ potential truncated at $r_{min}$ of 5 {\AA}.
For globular proteins \cite{JPCM} the commonly used value is 4 {\AA},
%as in ref. \cite{JPCM}, 
but the less constraining potentials appearing in our description
of the IDPs, and hence an enhanced flexibility of the chain,
require boosting the effective backbone stiffness by
making the excluded volume larger.\\

There are several conditions that need to be satisfied in order
to establish a contact. The first of these pertains to the distance,
$r_{on}$, at which the LJ potential between two beeds is
switched on in addition to the repulsion.
We set $r_{on}$ to be equal to $r_{min}$ for any given type of a contact.
The other conditions will be specified in the next two subsections.
An established contact is switched off if the $\alpha$-C$_i$--$\alpha$-C$_j$
distance exceeds 1.5$\;\sigma_{i,j}$.\\

\subsection{Effects of the directionality}

\begin{figure}[h]
 \centering
 \includegraphics[width=0.8\textwidth]{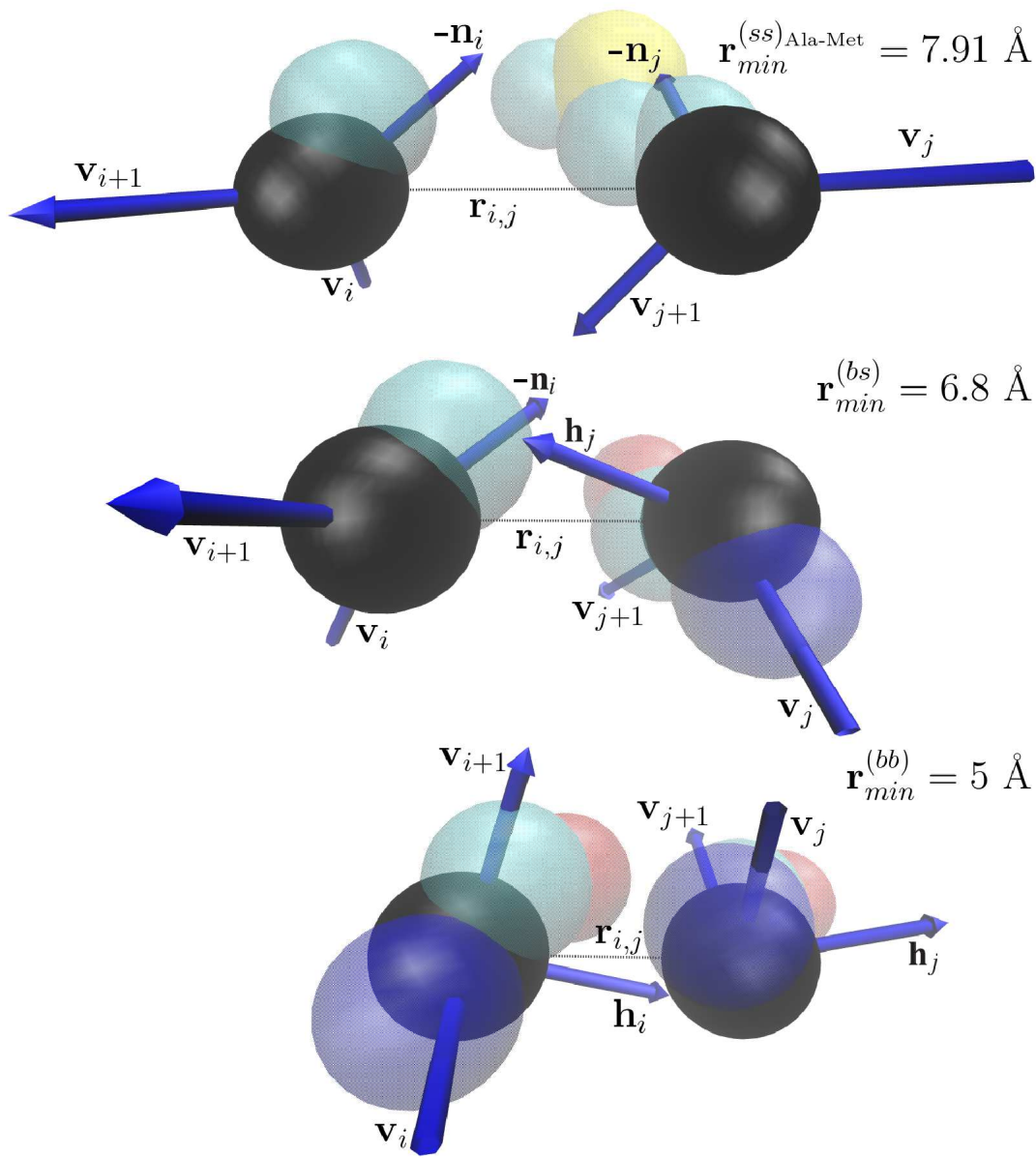}
 \caption{Examples of contact formation between Ala (the residue on the left) and Met.
The distance between the $\alpha$-C atoms (shown as spheres in black) are
7.91, 6.03 and 5.4 {\AA} top to bottom respectively.
The top, middle and bottom  panels illustrate formation of an ss, bs and bb
contacts respectively. 
The first two panels are obtained in one trajectory of an all-atom
simulation (performed using  NAMD). The bottom panel shows
a fragment of an $\alpha$-helix in the structure PDB:14GS.
The $\alpha$-C atoms are shown as black spheres.
%The other C atoms ($\beta$-C or backbone) are shown in green.
The other atoms are shown using the CPK coloring scheme.
The hydrogen atoms are not shown.
The sidechain atoms are represented by semi-transparent  spheres.
The arrows indicate either vectors ${\bf {h}}$ (for  the backbone-involving
interactions) or ${\bf {n}}$ (for the sidechain-involving interactions, as
indicated. Vector $r_{i,j}$ connecting $\alpha$-C atoms is drawn as a dotted line.
The threshold values of $r_{i,j}$, at which a given kind of contact
is considered to be setting in, are written on the right.
${\bf v}_i$ is defined as ${\bf r}_i - {\bf r}_{i-1}$.
 } 
 \label{arrows}
 \end{figure}

A bb contact may arise if the N-atom on the backbone part of
the $i$th residue can form the hydrogen bond with the O-atom on the
backbone part of residue $j$ or vice versa. This is allowed only if the
two atoms point towards each other. % and not if they point away.
This means that the vector from $\alpha$-C to O in one
residue should be approximately antiparallel to the
vector from $\alpha$-C to N in another. In addition, 
the lines set by these two vectors should nearly coincide
(instead of being far away). In order to capture this physics
by involving only the $\alpha$-C atoms, we use the
${\bf h}_i$ introduced in references \cite{tube,Rey} that starts from the $\alpha$-C
atom on residue $i$ and is perpendicular to the plane that is set
by sites $i-1$, $i$ and $i+1$, i.e. parallel to 
${\bf v}_i \times {\bf v}_{i+1}$, where ${\bf v}_i= {\bf r}_i -{\bf r}_{i-1}$
and ${\bf {r}}_{i,j}={\bf {r}}_{j}-{\bf {r}}_{i}$.
Examples of these vectors are shown on the bottom panel of Fig. \ref{arrows}. 
It can be seen that they should be lying along nearly parallel 
(or antiparallel, e.g. in antiparallel $\beta$-sheets) directions. 
This is captured by the three directional conditions imposed on the
formation of an bb contact \cite{Rey}:
a) $|cos({\bf {h}}_i,{\bf {r}}_{i,j})| > 0.92$ (the threshold angle of 23$^o$),
b) $|cos({\bf {h}}_j,{\bf{r}}_{i,j})| > 0.92$ and
c) $|cos({\bf {h}}_i,{\bf {h}}_j)| > 0.75$ (the treshold angle of 41$^o$).
The values of the thresholds have been obtained from statistical distributions 
of these angles \cite{Rey}.\\

Defining the directionality associated with the ss contacts
can be accomplished by introducing the normal vector \cite{Hoang}
\begin{equation}
{\bf n}_i = \frac{{\bf r}_{i-1}+{\bf r}_{i+1} - 2 {\bf r}_i}
{|{\bf r}_{i-1}+{\bf r}_{i+1} - 2 {\bf r}_i|}  \;\;\;.
\end{equation}
The negative normal direction ($-{\bf n}$) approximately points
from the $\alpha$-C to $\beta$-C atoms 
(see the top panel of Fig. \ref{arrows}).
A more accurate, but also more complicated, statistical expression 
for the location of the $\beta$-C atom can be
found in the references \cite{Covell,Micheletti}.  %refs. \cite{Covell,Micheletti}. 
However, this more accurate description  %given in ref. \cite{Micheletti} 
is valid for 82\% of the ss contacts determined by the OV criterion,
while the simpler description is valid for an even higher value of 97\%.
For an ss contact to form, it is necessary that the side chains point at each
other. There are two conditions for this to happen \cite{Hoang}:
1) $\cos({\bf n}_i,{\bf r}_{i,j})<0.5$ and %(the angle of 60$^o$)
2) $\cos({\bf n}_j,{\bf r}_{j,i})<0.5$ (the threshold angle of $60^o$).
The conditions for the ss contact are qualitatively distinct
from those for the bb contact because the nature of the ss contacts
is much more diverse (resulting also in less restrictive thresholds),
whereas the bb contacts have a specific hydrogen bonding pattern \cite{Rey}.
\\

Similar considerations lead to the following two conditions
for a bs contact to form 
(the sidechain of the $i$th residue interacts with the backbone of the $j$th residue, see the middle of Fig. \ref{arrows}):
1) $\cos({\bf n}_i,{\bf r}_{i,j})<0.5$ and 
2) $|\cos({\bf h}_j,{\bf r}_{j,i})|>0.92$.\\

As demonstrated in Fig. \ref{distri}, most of the OV-based contacts
in the native states of the structured proteins satisfy the directional
criteria as listed above.
Exceptions occur mainly in the bs contacts, where the nature of the contact 
may not be fully captured by directional criteria associated separately with the ss and bb cases.
The number of contacts formed between the backbone and hydrophobic sidechains is comparable to
those formed with polar sidechains. Thus the nature of the bs overlap contacts can be difficult to 
describe and the corresponding characteristic distances are much less 
specific than those corresponding to the ss contacts.
Imposing directional criteria shifts $r_{min}$ of the bs contacts from 6.8 \AA to 5.4 \AA,
 but this does not change the results in any significant manner.\\

Acceptable values of $\cos({\bf n}_i,{\bf r}_{i,j})$ seem to be largely independent
from the $r_{i,j}$ distance itself. Examples of distributions on the plane corresponding 
to  $\cos({\bf n}_i,{\bf r}_{i,j})$ and $r_{i,j}$ are shown in Fig. S4 in the
Electronic Supplementary Information (ESI).
We do not introduce any potentials that would favor the directionalities
to stay within the appropriate bounds -- 
they are checked only at the instant when the contact is established.\\

\subsection{The types and specifity of effective residues}

We group the residues into six classes: 1) Gly, 2) Pro, 3) hydrophobic:
Ala, Cys, Val, Ile, Leu, Met, Phe, Tyr, Trp, 4) polar: Gln, Ser, Thr, Asn, His
5) negatively charged: Asp, Glu, and 6) positively charged: Arg, Lys
(see Table \ref{aaproperties}). The
last two classes count as polar but they also carry charge. The distinction 
between the polar and hydrophobic residues matches (with the exception of 
Ala) the clustering (based on the eigenvalue decomposition) of the
Miyazawa-Jernigan interaction matrix into two clusters \cite{alphabet}. Contacts 
of the ss kind can be made between the hydrophobic residues \cite{Hoang}, 
but also between the polar ones (crucial for modeling of polyQ)
and between the polar and hydrophobic residues.
The unlike-charged polar residues may form salt
bridges but the like-charged ones merely repel.
The energy parameter for each type of the ss contacts is the same.
This, together with allowing for the formation of hydrophobic-polar ss contacts, 
distinguishes our model from the simple ,,HP'' models\cite{Hoang}, where such contacts 
either have reduced strength or are not allowed to arise. Thus, 
even though our model cannot capture the details of all possible interactions
(for instance, between the $\pi$ electrons on the faces of aromatic rings
and cations), our models allows for them statistically.
%taking their directions and distances in a statistical way.}
\\

Real cysteine residues may form disulfide bonds, but this process is not covered 
in this paper, as none of the considered systems %, except polyC, 
seems to form such bonds. They can be implemented in a coarse-grained model
as discussed in references \cite{disulMorse,Thirumalai}.\\
%refs. \cite{disulMorse,Thirumalai}.\\

Each residue is allowed to form up to $z_s$ contacts with other residues.
The number $z_s$ is residue-dependent. We take $z_s=n_{b}+\min(s,n_{H}+n_{P})$, 
where $n_{b}$ is the 
permitted number of backbone contacts, $s$ is the maximum allowed number of 
the sidechain contacts and $n_{H}$, $n_{P}$ are the maximal numbers of sidechain 
contacts with hydrophobic and polar sidechains of other residues, respectively.
The values of these parameters are listed in Table \ref{aaproperties}.
For a bs contact, one backbone "slot"
assigned to one residue must be combined with one sidechain "slot" assigned
to the other residue. The backbone does not count as polar in bs contacts, 
so it affects only the total $s$ limit for a given sidechain. 
Treating the backbone as a polar entity merely changes the proportions 
between the types of contacts arising without improving the results. 
$n_{b}$ is equal to 2,
corresponding to two possible hydrogen bonds.
In the case of Pro, $n_{b}=1$.\\

The numbers $n_{H}$, $n_{P}$ and $s$ depend on the type of the residue and
were derived from the structure database by calculating, for each residue of a given 
type, the distribution of its observed numbers, $s_c$, of the ss contacts, i.e.
the coordination numbers. $s$ was operationally defined
in relation to $s_{max}$ - the values of $s_c$ corresponding to
the maxima in such distributions. For the hydrophobic residues, the
distributions are broad and we take $s=s_{max}$. For the polar 
residues, the distributions are narrow and then we take $s=s_{max}+1$.
The polar residues usually stay near the surface of a protein
and thus some of their possible 'contacts' are
with the molecules of water instead of with the other residues
and hence the under-count in the distribution.
This is also the reason why we do not determine $s$
by counting the possible hydrogen bonds that could arise.
The values of $n_{H}$, $n_{P}$ were determined similarly, by calculating two-dimensional
distributions of pairs of numbers of hydrophobic and polar contacts (with maxima at points
$s_{Hmax},s_{Pmax}$). Then, $n_{H}$ corresponds to
the maximum $s_{Hmax}$ of such distribution (the most common number of hydrophobic ss contacts for a 
given type of amino acid), $n_{P}$ to $s_{Pmax}+1$.
The values of $s$, $n_{H}$, $n_{P}$ for the different types of residues 
are listed in Table \ref{aaproperties}.

It has been suggested that electrostatics, specifically the net charge per 
residue \cite{Mao}, may influence the nature of conformational ensembles
of the IDPs substantially. Thus,
in addition to the ss, bb, and bs contacs, the
charged residues interact with each other via the modified 
Debye-Hueckel potential \cite{Debye} 
\begin{equation}
V_{D-H}=\frac{e^2\exp{(-r/\lambda)}}{4\pi \varkappa \varkappa_0 r}
\end{equation}
with the screening length $\lambda=10$ {\AA}.
Following Tozzini {\it et al.} \cite{Tozzini}, the relative permittivity is
taken to be $\varkappa=4\;\textup{\AA}^{-1}\; r$ (generally, $\varkappa$ changes with the distance sigmoidally so
that it is around 4 well within the protein and around 80 at
large distances; $\varkappa _0$ is the permittivity constant). 
This leads to the effective electrostatic potential which is approximately equal 
to $V_{el}(r)=\frac{85 \exp{(-r/\lambda)}\;\epsilon\textup{\AA}^2}{r^2}$. 
This long-range interaction does not count into the $z_s$ limit.
When two oppositely charged residues form an ss contact, 
their interaction becomes described by the standard LJ contact potential
with $r_{min}$ taken from Table \ref{rmin}, in an analogy to the way the salt bridges
are described in the structure-based models of the structured proteins.
When a salt bridge is adiabatically turned on, other electrostatic interactions
made by these two residues are turned off in the same manner, 
so that the local geometry of the bridge is stable.
However, we have found that this precaution does not affect the results
noticeably.
Like-charged residues cannot form the ss contacts but they may participate 
in the bb contacts. This is possible for certain conformational geometries
or for situations in which at least one of the charges is effectively
removed due to the formation of the ionic bridge with another residue.
The first and last residues in the chain 
are not allowed to form attractive inter-terminal LJ contacts
so that the protein does not form a closed loop % in ref. \cite{Rey},
(in Enciso and Rey's model \cite{Rey} the inter-terminal contacts may exist, but 
their strength was reduced).
In an all-atom model, the termini do come with opposite electric charges
but this is usually not so in coarse-grained models in which the terminal
residua have the same properties as the non-terminal ones.

\subsection{The backbone stiffness}

The backbone stiffness is usually defined in terms of bond and 
dihedral angles. In structure-based models of  proteins, 
these angles can be measured as deviations from their local
native values.  In the IDP case there are no such reference
values and we resort to a statistical approach as proposed by
Ghavani {\it et al.} \cite{Ghavani}\\

In this approach, and when determining the distribution of bond angles,
one divides the residues into three types: Gly, Pro and X, where
X denotes all other 18 residues. There are 27 possible combinations
of these residues and the bond angle associated with the middle of
the three consecutive residues is governed by a distribution
specific to the triplet.  The distributions have been obtained by
matching to the experimental Ramachandran angles \cite{Ghavani}
and are found to be fairly broad: they typically span about 60$^o$
whereas the structure-based potentials are significantly narrower.
We use the potentials that distinguish only the middle residue and whether
it precedes proline, which results in dealing with only 6 possible combinations.
The sequential order of the residues matters because it defines the local
chirality. Thus we can only distinguish cases in which a proline
is preceded but not when it is succeeded.
The examples of statistical potentials are shown in Fig. S5 in ESI.
In our molecular dynamics simulations, we fit each statistical potential to 
a sixth degree polynomial with the coefficients listed in Table S1 in ESI.\\

The dihedral angle requires a consideration of four consecutive residues
that can be viewed as a superposition of two triplets that overlap
at two central residues. In the three-alphabet case, there are 9 possible 
overlaps and thus 9 distributions. We show examples of the statistical potentials for these
distributions in Fig. S6 in ESI.
We fit the statistical potentials to the formula:
$a \sin( x )+b \cos( x )+c \sin^2( x )+d \cos^2( x )+e \sin( x ) \cos( x )$,
where the values of the coefficients are also listed in Table S2 in ESI.\\

Both of the bond-angle and the dihedral potentials 
were obtained by the inverse Boltzmann method
at the room temperature applied to the random coil database statistics. 
These potentials correctly describe properties of denaturated proteins \cite{Ghavani}, 
that can be assumed to have no contacts.
A priori, the energy scale involved may not exactly match the one associated
with the contact energy. We have checked, however, that doubling the
amplitudes of the angle-involving potentials does not affect the
geometrical parameters of the test systems in any significant
manner.

\subsection{The molecular dynamics simulations}

The solvent is implicit.
The time evolution is defined in terms of
molecular dynamics with  the velocity-dependent
damping and the Langevin noise, both representing
the influence of the solvent. The noise corresponds
to temperature $T$.
The characteristic timescale, $\tau$ is of order 1 ns since
the system is considered to be overdamped so that the motion of the
beads is diffusional instead of ballistic. 
The damping constant, $\gamma$, is taken to be
$2m/\tau$, where $m$ is the typical mass of a residue. More realistic
values of $\gamma$ should be about 25 times larger \cite{Veitshans}
but adopting them would lead to much longer conformational dynamic.

As described earlier, there are three types of requirements
for a contact to be legitimate: 1) on the distance between the residues,
2) on the proper directional placement of the implicit side groups
and backbone hydrogen bonds, 3) on the number of contacts that the
residues are permitted to form. If all of these requirements
are fulfilled, a contact is made. A sudden creation of a contact
may lead to instabilities and hence we switch the corresponding
contact potential on adiabatically: the depth of the potential
well increases linearly form 0 to $\epsilon$ within the timescale of 10$\;\tau$.
This timescale is sufficiently long for the system to 
thermalize so the process is quasi-adiabatic.
We have observed that shorter switching-on times lead to heating of the system
(see the top panel of Fig. S7 in ESI). 
Longer times, up to 40 $\tau$, do not affect the statistical properties that
we measure. Still longer times make the chains more mobile which shows as a
slight expansion of the system (bottom panel of Fig. S7 in ESI).
Increasing $\gamma$ is expected
to lead to more stable dynamics and longer permissible switching-on times. \\
 
We integrate the equations of motion
by the fifth order predictor-corrector algorithm \cite{Tildesley}
and each $\tau$ is discretized into 200 steps. Thus the
adiabatic switching-on process is accomplished in 2 000 steps. Switching off
of the contacts, when the contact distance crosses 1.5$\;\sigma_{i,j}$,
is implemented with the same time scale. If a pair of residues
is connected by a contact of one type, it cannot be simultaneously
connected by a contact of another type.\\

As explained in reference \cite{wolek2} %ref. \cite{wolek2} 
the kinetic and thermodynamic
behavior of a coarse-grained model depends on the type
of the backbone stiffness used. The $T$-range in which folding
of structured proteins
is optimal is around 0.3 -- 0.35$\; \epsilon /k_B$ if the
backbone stiffness is described by a chirality potential. In this
case then 0.35\;$\epsilon /k_B$  corresponds to the room temperature, $T_r$,
which is consistent with the callibrated value of $\epsilon$. However,
for the bond and dihedral potential $T_r$ shifts to about 
0.7$\;\epsilon /k_B$, because these terms contribute more substantially
to the energy. \\

In order to determine $T_r$ for our IDP model, we perform
a similar folding test for several globular proteins with
the PDB structual codes of 1GB1, 1TIT, 1UBQ, and 2M7D.
We take the native contacts map but adopt the statistical
angular potentials for the backbone stiffness.
We find that the kinetic optimality is reached around
0.3 -- 0.35$\; \epsilon /k_B$. 
In addition, Fig. S8 in ESI shows that for 0.3$\;\epsilon /k_B$,
the average end-to-end distance matches the room-temperature all-atom results
for polyQ for $n$ up to 60 and it does not for higher temperatures.
The matching at 0.2$\;\epsilon /k_B$ is comparable, but folding
is significantly worse than at 0.3$\;\epsilon /k_B$
due to the emergence of more potent kinetic traps.
Moreover, Fig. S8 demonstrates a similar matching to the
experimental results for the flanked versions of polyQ.
We have also found that best agreement with experimental data for 
polyproline occurs for temperatures in range 0.3 -- 0.4$\;\epsilon /k_B$. 
(results for $T=0.3\;\epsilon /k_B$  will be discussed later). %are shown in Fig. \ref{exper1})
Thus, in our simulations we take
0.3$\; \epsilon /k_B$ as the $T$ for which most simulations were made.\\

The results on the geometry of the systems were obtained based on 100
independent trajectories, each lasting for 100 000 $\tau$. The time evolution in the
first 5 000 $\tau$  was excluded from the data acquisition to allow for equilibration. 
The conformations were saved every 100 $\tau$, unless stated otherwise. 
For structured proteins, we started from the native structure whereas for the 
IDPs -- from a self-avoiding random walk.

\section{Results}

The main purpose of the performed simulations is to provide
a validation of our model.
We use three parameters to characterize the geometrical properties
of an IDP. One is the time-averaged 
radius of gyration $R_g=\sqrt{\langle r_{g}^2 \rangle}$,
where $r_g$ is the instantaneous value of this radius.
Another is $l$ -- the time averaged end-to-end instantaneous distance $d_{ee}$
($l = \langle d_{ee} \rangle$). The third is
$\sigma = \sqrt{l^2 - \langle d_{ee} \rangle^2}$, which is 
the dispersion in $d_{ee}$.
As described by R{\'o}{\.z}ycki {\it et al.} \cite{linkers,linkers0} 
%Following refs. \cite{linkers0,linkers}, 
$\sigma$ can be used to define the effective elastic stiffness,
$\kappa$, of the system, since energy equipartition implies that
$\frac{1}{2}\kappa \sigma ^2 =\frac{3}{2} k_BT$.
The nature of our coarse-grained approximation does not allow
us to make predictions about local structural parameters
such as the Ramachandran angles and thus to make the
corresponding comparisons.\\

Generally, we test the predictions of our model against the results
obtained through all-atom simulations, in which the geometrical
parameters were calculated, and against the experimental results.
We have chosen the NMR and SAXS experiments in which few
interpretational assumptions about the parameters were made.
The comparisons are shown in Figs. \ref{polyQ}, \ref{theor1}, and \ref{exper1}.
The first two of these present comparisons to the all-atom results
and the last -- to the experimental ones.\\

\subsection{PolyQ and polyV}

\begin{figure}[h]
 \centering
 \includegraphics[width=0.5\textwidth]{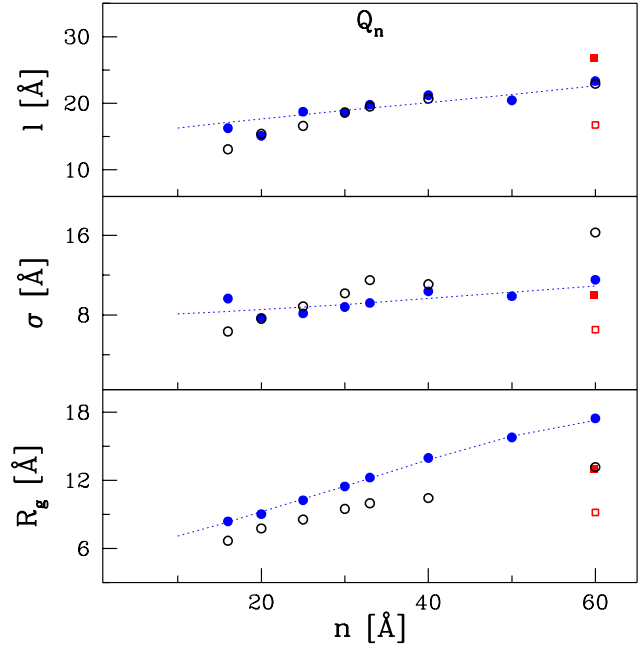}
 \caption{The results obtained
by our IDP coarse-grained model for Q$_n$ (solid blue circles) and V$_n$
(solid red squares)  as a function of
the number of residues, $n$. The top panel is for $l$, the middle for
$\sigma$, and the bottom one for $R_g$. The open circles correspond to the
results obtained by all-atom simulations \cite{Gomez} for Q$_n$. %in ref. \cite{Gomez}. 
The open red squares -- only for $n$=60 -- correspond to the results
obtained by all-atom simulations \cite{Cossio} for V$_{60}$.% in ref. \cite{Cossio}.
The lines are guides to the eye. The sizes of the symbols are of the
order of the error of the mean.
 }
 \label{polyQ}
 \end{figure}

Fig. \ref{polyQ} pertains to polyQ and polyV for $n$ of up to 60.
We use the notation Q$_n$, V$_n$, and similarly for other
amino-acid homopolymers. Cossio {\it et al.} \cite{Cossio} have generated
30 063 statistically independent conformations for V$_{60}$
by using all-atom simulations in a meta-dynamics approach involving
the replica-exchange method. The solvent was implicit and treated
within the generalized Born surface area approach. They
concluded that only a fraction of these conformations is structurally similar
to the proteins listed in the CATH database. They took it as an
evidence that there are evolutionary pressures that favor only selected
classes of conformations. Here, we calculate $l$, $\sigma$, and $R_g$ for
7 077 confromations of V$_{60}$, that were made available to us by the
authors, by assuming that these conformations approximately
represent an equilibrium trajectory. Our model is
seen to generate parameters that are larger by a factor of around 2
(depending on which quantity is considered) for $n$=60, but the differences
are expected to get smaller for lower values of $n$. \\

It should be noted that
V$_n$ does not exist in nature whereas long tracts of polyQ do arise
in various proteins, for instance in the context of Huntington disease.
G\'omez-Sicilia {\it et al.} \cite{Gomez} have applied the approach of
%ref. \cite{Cossio} to Q$_n$ with $n$ between 16 and 80. 
Cossio {\it et al.}\cite{Cossio} to Q$_n$ with $n$ between 16 and 80. 
The statistics of the independent conformations were 
308, 491, 330, 422, 479, 322, 269, 246, and 108 for
$n$=16, 20, 25, 30, 33, 38, 40, 60, and 80 respectively. We do not consider
$n$=80 here due to the smaller statistics.\\

Fig. \ref{polyQ}
demonstrates that the geometrical parameters,
calculated based on the statisticaly independent conformations
obtained by G\'omez-Sicilia {\it et al.}\cite{Gomez} % in ref. \cite{Gomez} 
agree fairly well with the results
of our model, especially in the case of the parameter $l$.
Generally, the smaller the $n$, the closer the 
agreement, especially in the case of $l$. We consider the agreement
to be adequate enough to model gluten (under study) which
consists of many Q-rich chains \cite{gluten1}.
We have observed that a better agreement with the all-atom data on polyQ
can be obtained by increasing the value of the parameter $s$ 
(the number of possible sidechain contacts) from 2 to 3.
However, we keep $s=2$ in order to achieve optimal consistency
for all systems considered here.
Increasing the limit for backbone contacts, $n_b$, from 2 to 3, results
in a similar improvement for $R_g$ (though not for $\sigma$) for this system.
\\

One of the findings of G\'omez-Sicilia {\it et al.} \cite{Gomez} was %ref. \cite{Gomez} was  
that a substantial fraction (9.3\%) of the statistically independent
conformations of Q$_{60}$ are knotted and that the sequential extension
of the knotted segments starts at about 36 residues. Such knotted conformations
may jam the proteasome and thus lead to toxicity \cite{proteasome}.
In experimental systems\cite{Petruska}, the toxicity arises above a threshold of
about 35.% as demonstrated in ref. \cite{Petruska}. 
Our coarse-grained model has produced several knotted conformations 
for Q$_{60}$ (all with the shallow knots) and
none for V$_{60}$. In the all-atom model, V$_{60}$ had a factor
of 3 smaller propensity to form knots than Q$_{60}$ so there is
a qualitative agreement.
An example of a knotted conformation obtained for Q$_{40}$ is shown in
the rightmost panel of Fig. \ref{aconf}. 
Increasing $s$ from 2 to 3
has been found to enhance the probability of knotting (but still below 9\%).
(We find that more substantial knotting propensities are exhibited by polyW).\\

Since the IDP systems flow from one conformation to another, it is interesting 
to ask for how long do they stay in particular conformations.
One measure to assess this is to probe the system every 1$\;\tau$
and determine the time needed to deviate in RMSD (root mean square deviation)
from the considered conformations by 5 {\AA}. For Q$_{60}$ the distribution
of such times extends up to 65$\;\tau$ and has the mean of 24$\;\tau$
and the dispersion of 10$\;\tau$.
A typical duration time obtained in the all-atom simulations \cite{Gomez}
is up to of order 20 ns, i.e. comparable to that obtained in the coarse-grained simulations.\\

\begin{figure}[!h]
\centering
\includegraphics[width=0.6\textwidth]{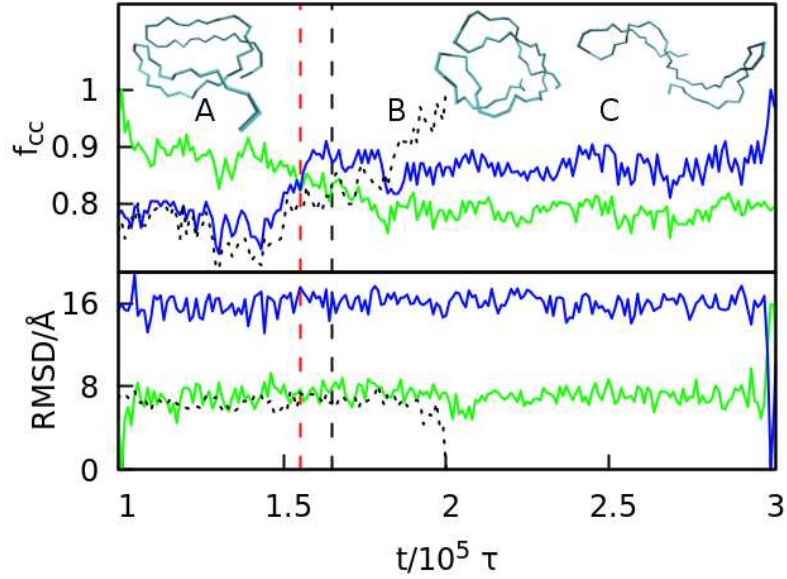}
\caption{The time-dependence of $f_{cc}$ (top) and RMSD (bottom) for Q$_{60}$.
It is calculated every 1000 $\tau$ with respect to three reference 
structures obtained at  $10^5~\tau$ (A), $2\cdot 10^5~\tau$ (B, dotted line), 
and $3\cdot 10^5~\tau$ (C) in an example of a trajectory.
The reference structures are depicted at the top.
The red vertical line indicates a transition from a similarity in $f_{cc}$
between A and C. The black vertical line: between A and B.}
\label{rmscnt}
\end{figure}

Another way to characterize conformational transformations is to use
the fraction, $f_{cc}$, of the contacts that are common with a reference
structure. The RMSD is more related to the overall shape whereas $f_{cc}$
to the nature of the local cohesion. For instance, bending of the chain
segments results in an increase in the RMSD but may leave $f_{cc}$ largely unchanged,
indicating persistence of the local structure.
The time dependence of the two measures for a trajectory obtained for
Q$_{60}$ is shown in Fig. \ref{rmscnt}. We consider three reference structures:
one (the lines in green; denoted by A) obtained at $10^5 \;\tau$,   
another (the lines in black; denoted by B) at $2\;10^5\;\tau$ and still another
(the lines in blue) at $3\;10^5 \;\tau$. The red vertical line indicates
time at which, based on $f_{cc}$, the conformation stops being more
alike to conformation A and becomes more alike to conformation C. This happens at
about $1.53\;10^5 \; \tau$ indicating that the changes in the contact
map take place in the scale of microseconds. Interestingly, a similar transition
to conformations similar to B (the black vertical line), which is closer in time, takes place
at the later instant of $1.67\;10^5\; \tau$.
In contrast, the RMSD crosses the 5-{\AA} treshold much more rapidly -- within
$0.05\; 10^5\;\tau$.
A more detailed characterization of the time evolution in terms of $f_{cc}$
can be found in ESI (Fig. S9).\\

Experimental FRET data for polyglutamine chains of length $m$ equal to 8, 12, 16, 20 and 24
have been obtained by Walters and Murphy \cite{polyQlen} for two values of pH: 7 and 12.
These chains, however, are flanked by the sequence KKW on the N-terminus and by AKK 
on C-terminus so the backbone length $n=m+6$.
Thus the model-based results require recalculations, especially because
the presence of the lysines brings in electrostatic effects, 
providing another kind of test of the model.
Fig. \ref{theor1} shows our results compared to the experimental ones for
pH=7. The systems are denoted as KQmpH7 and the data points are shown in green.
There is a close agreement for $l$ but a discrepancy  for $\sigma$.
The experimental results on $\sigma$ apper unreasonably small though,
particularly at $n$ of 30 (the case shown in the figure), where the Coulomb 
effects should be too weak to keep
the system stiff and barely fluctuating.
All theoretical data points are distinct from the range of values found in
systems without the flanking lysine (Fig. \ref{polyQ}). If we assume that
changing pH from 7 to 12 affects mostly the protonation level of
the lysine (according to Enciso {\it et al.} \cite{pHEnciso}), then this
should be equivalent to replacing the charged lysine by the neutral asparagine.
Our theoretical results with the asparagine show correlations
with the experimental data at pH=12 (see Fig. S10 in ESI) of a similar quality as with
the lysine at pH=7.\\

\subsection{PolyA and polyP}

\begin{figure}[!h]
\centering
\includegraphics[width=0.8\textwidth]{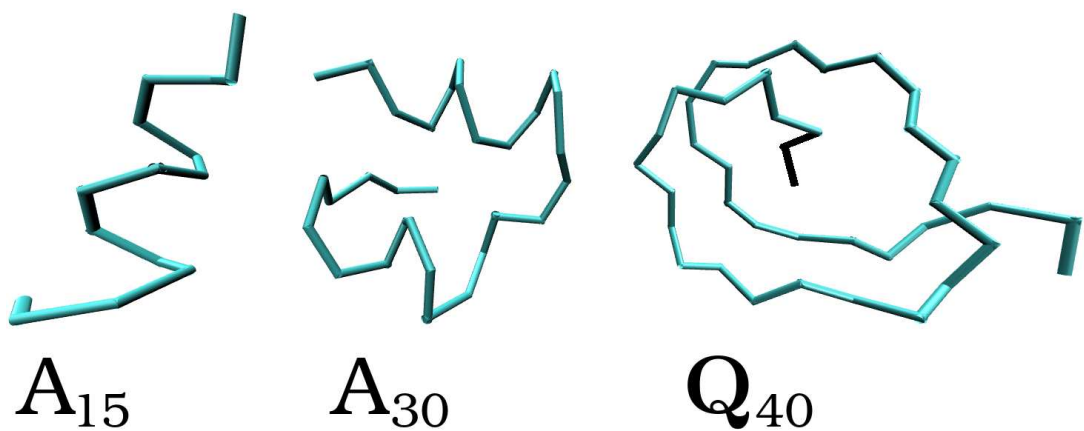}
\caption{Examples of conformations of three homopeptides: A$_{15}$ (left), 
A$_{30}$ (middle) and Q$_{40}$. The latter conformation is knotted.
The N-terminal end of Q$_{40}$ is shown in black. It crosses the loop inwards,
forming a shallow knot.}
\label{aconf}
\end{figure}

Another interesting system is polyA. The all-atom results, which used the TIP3P model
for the molecules of water \cite{TIP3P} and the CHARMM36 protein force field, have
been derived \cite{linkers} %in ref. \cite{linkers} 
for $l$ and $\sigma$ and for a set of values of $n$. They are compared
to our coarse-grained results in Fig. \ref{theor1}. We observe that there is
a good agreement for small values of $n$, up to 15, 
when the conformations are helix-like (see also ref. \cite{Cossio})
as illustrated in the left panel of Fig. \ref{aconf}.
At larger values of $n$,  the coarse-grained data
fall down relative to the all-atom ones. We attribute the discrepancy to the
differences in the timescales. In the all-atom simulations it is 100 ns
whereas in the coarse-grained model -- of order 5 $\mu$s, {\it i.e.}
50 times longer and repeated in 100 trajectories. In the latter situation,
the chain can explore significantly more phase space and adopt bent
conformations such as the one shown in the middle panel of Fig. \ref{aconf}. Such
bent conformations allow for a closer approach of the termini to each other than
in the case of a helix. They have been observed in simulations with an implicit 
solvent \cite{Bleha}
%reported in ref. \cite{Bleha} 
for $n$ exceeding 40.
The threshold $n$ above which the bent conformations may appear must depend on 
the description of the backbone stiffness -- in our case the backbone is
rather soft (the constraints on the bond angle are weak) and hence
the lowering of the threshold. A better modeling of the bond-angle potential
should increase the treshold. \\

The results for polyP are shown in Fig. \ref{exper1}. The agreement with the 
FRET-based data \cite{polyP} on $l$ and $\sigma$ is pretty good.
All three geometrical parameters exhibit a relative steep growth 
with $n$, reflecting the fact that the shape of the conformations
is governed primarily by the backbone-stiffness potentials so that the details
of any non-local potentials have only a minor impact.\\

\subsection{Other systems}

\begin{figure}[h]
 \centering
 \includegraphics[width=0.5\textwidth]{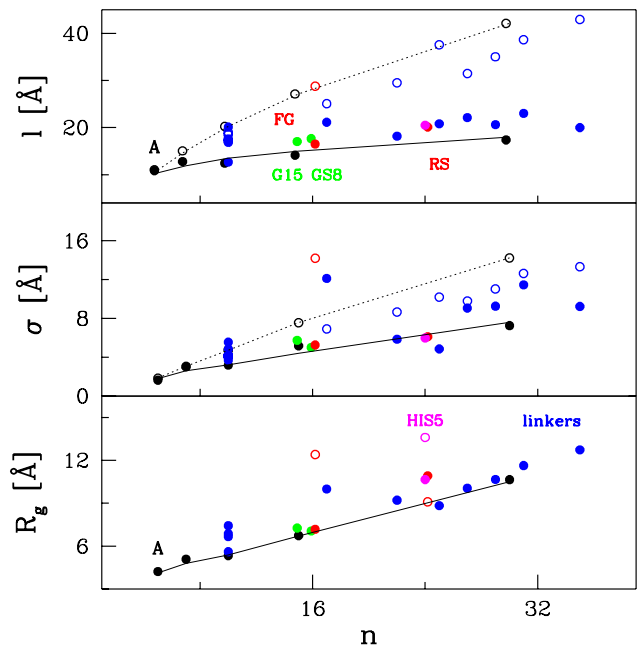}
 \caption{Similar to Fig. \ref{polyQ} but for other indicated systems.
The results obtained
by our IDP coarse-grained model are shown using the full circles
and those obtained by all-atom simulations -- by the open circles.
The black data points are for polyalanine. The blue data points --
for selected linkers\cite{linkers}. The selected linkers are
those with $n=10$ and $n>15$. The red data points are for the systems
denoted as FG and RS. Note that in the case of FG, the all-atom results
are available only for $R_g$. The green data points are for G15 and GS8.
The magenta data points are for His5.  }
 \label{theor1}
 \end{figure}

\begin{figure}[h]
 \centering
 \includegraphics[width=0.5\textwidth]{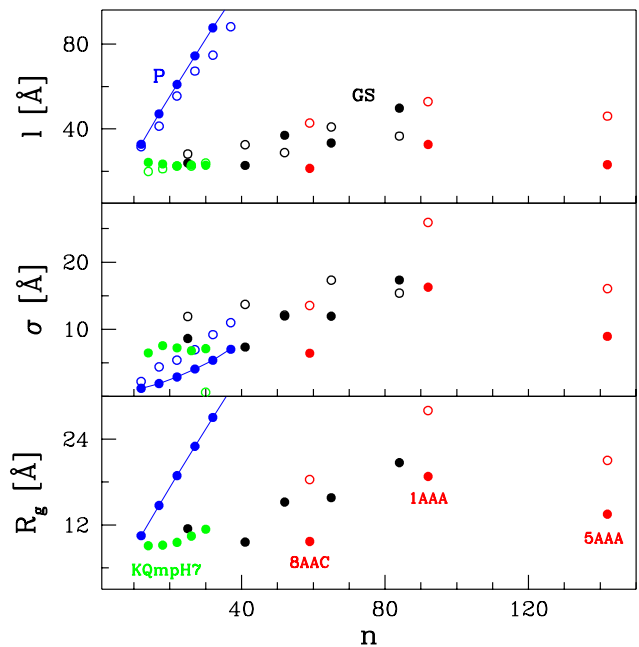}
 \caption{Similar to Figs. \ref{polyQ} and \ref{theor1}
but now the comparison is to the results obtained experimentally
(the open circles).
The blue data points are for polyproline. 
The black data points are for "GS".
The red data points are for the three indicated systems
from PEDb. 6AAA and 9AAA disagree with the theoretical findings
more significantly and are not shown.
The green data points are for KQn at pH=7.  }
 \label{exper1}
 \end{figure}

In studies on glycine-rich systems \cite{glyser,GGS}
%reported in ref. \cite{glyser} and \cite{GGS},
the FRET efficiency is not used to calculate $l$ directly. Instead, it serves as a
constraint that is employed to reconstruct the distance distributions by using simple models. 
These systems incorporate repetitive sequences containing glycine 
(like GGS or GGGGS, collectively named ,,GS''). They are expected to be prone to a
substantial flexibility. This expectation is confirmed (see the plot of $\sigma$ in
Fig. \ref{theor1}). Furthermore, there is a good agreement of the  model
results with the FRET-based data.\\

We now consider the disordered cellulosomal linkers that are listed in Table 1
of reference \cite{linkers}. In order to avoid crowding in Fig. \ref{theor1},
we show the results only for the linkers with $n$=10 and $n\ge 17$ (the symbols in blue).
The data points are scattered -- however, they are comparable 
to the all-atom explicit-solvent simulations for these systems (there are no
all-atom results for $R_g$).\\

IDP's may require using different force fields than the structured proteins.
Huang {\it et al.} \cite{Nawrocki} have recently proposed a new force field,
CHARMM36m, which should show an improved performance both in the
IDP and structured cases.
Several proteins were considered as test cases. Among them, the FG-nucleoporin
and the RS peptide, both disordered. Fig. \ref{theor1} shows 
that our model gives results that are %compatible with 
comparable to those obtained by the novel force field  %using CHARMM36m 
($l$ and $\sigma$ are available for the RS peptides and $R_g$ for the
FG-nucleoporin peptide).
The agreement gets worse for Histatin-5 (His5,
used to parameterize the Amber and Gromos force fields \cite{his5})
probably because we consider His to be uncharged. An admixture 
of the charged states of His should introduce electrostatic repulsion and thus
larger values of $R_g$.\\

The Protein Ensemble Databank (PE-DB) \cite{pedb} contains a number of structural
data which were obtained through simulations which are considered to be
consistent with the ensembles obtained in the NMR and SAXS experiments
data on IDP's. These proteins have $n$ considerably larger than
the systems considered so far.
We find (see Fig. \ref{exper1}) that our results on $R_g$,
though not coinciding with the PE-DB results, are of comparable size.
However, the discrepancies are larger for 6AAA and 9AAA (not shown).
It should be noted, however, that the PE-DB results we compare to were obtained 
with the use of particular all-atom and/or coarse-grained models. In addition
they were derived  for a given
timescale which may or may not be adequate.\\

\subsection{Comparisons of homopeptides}      % of length 30}

\begin{figure}[!h]
\centering
\includegraphics[width=0.43\textwidth]{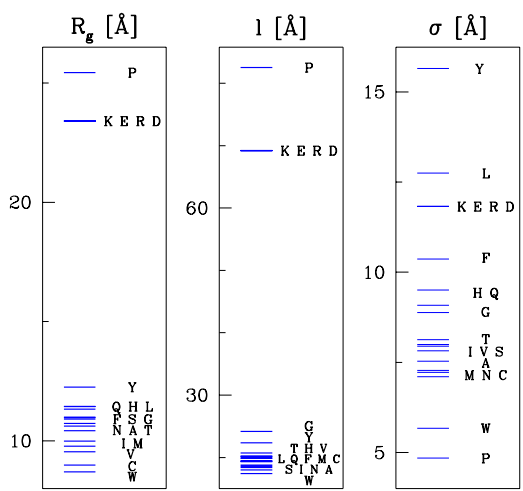}
\caption{The geometrical parameters $R_g$, $l$ and $\sigma$  
for the homopeptides of length 30. The error of the mean does not exceed 1, 3 and 2\% for 
$R_g$, $l$ and $\sigma$ respectively.
The types of the amino acids are
indicated in the single letter convention. If we considered the histidine 
residues to be charged, the parameters would be the same as for K,E,R and D.}
\label{geom30}
\end{figure}

\begin{figure}[!h]
\centering
\includegraphics[width=0.43\textwidth]{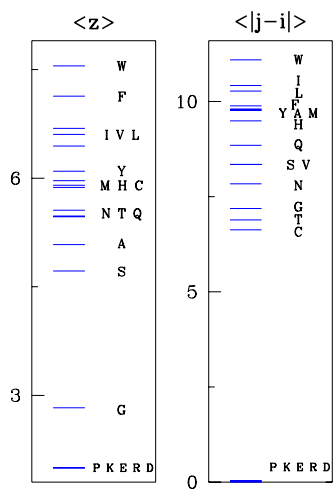}
\caption{Similar to Fig. \ref{geom30} but for the average coordination
number and the average sequential distance in the contacts. Contact order
is this distance divided by $n$. The error of the mean does not exceed 0.5 and 3.3\% 
for $<z>$ and $<|j-i|>$ respectively.}
\label{con30}
\end{figure}

Table \ref{aaproperties} lists the known maximal lengths of
tracts formed by amino acids of the same type, as obtained
from the UniProt database \cite{UniProt}. The longest of them
correspond to Gln (79), Ser (58),  and Asn (46). For Ala the
length is 24 but for the sligthly larger Val -- only 9.
Nevertheless, it is instructive to simulate homopeptides
of unrestricted length. In particular, Cossio {\it et al.}
have studied the chains of Val with $n$=60. Here, we simulate
all homopeptides of length 30 to bring out the differences
in behavior between them, as derived using our model. The results
on the geometrical parameters are shown in Fig. \ref{geom30} and on the
properties of the contacts in Fig. \ref{con30}. The properties
of the contacts are assessed by providing the average coordination 
number, $<z>$, and the average sequential distance in the contacs, $<|j-i|>$.
The former includes the contacts made by the dipeptide bonds along the backbone.
The latter, when divided by $n$, defines a parameter known as the contact order.\\

All of the homopeptides of length 30 appear to be IDP's since $\sigma$
is always substantial: it exceeds 5 {\AA}. In agreement with Uversky \cite{Uversky1}
and Mao {\it et al.} \cite{Mao},
there is a clear difference between the charged polymers (polyK, polyE, polyR, and
polyD) and the collapsed non-charged systems.
The charged homopetides have the largest values of $R_g$ and $l$ as they do not
form any contacts. PolyP comes next -- these chains also do not form contacts.
Due to the restricted bending capabilities, polyP is the stiffest of
the homopolymers considered here (the lowest vakues of $\sigma$).
However, the properties of polyW appear to be the most extreme: it shows
the lowest values of $R_g$ and $l$ and 
the highest values of $<z>$ and $<|j-i|>$. These features result from the
largest value of $r_{min}$ and hence from the  dominance 
of non-local ss contacts (every W residue can form up to 4 of them). 
There are almost no 
bb contacts in the simulated polyW systems. PolyW is also the most likely
to generate knotted structures. The knots appear in
16\% of the trajectories and last for about 5\% of total simulation time.\\

Other hydrophobic homopolymers are generally more compact and less 
flexible than the polar ones, with the exception of polyG and polyA
that are endowed with short side chains. This can explain the fact that
the longest hydrophobic residue tracks seen in the UniProt database are shorter 
than those made from the polar residues (again with the exception 
of polyG and polyA that are more flexible and thus more accomodating.)\\

Another way to compare the homopeptides is to study the scaling behavior of
radius of gyration in a broad range of the values of $n$: between 10 and 100.
For homopolymers with $n \neq 30$ each MD trajectory was 50 000 $\tau$ long.
We observe a good agreement with the power law for
\begin{equation}
R_g \sim n^{\nu}
\end{equation}
The values of the exponent $\nu$ are listed in Table \ref{aaproperties}.
We find that $\nu$'s for the large hydrophobic amino acids are smaller 
(which corresponds to the poor solvent regime \cite{Flory}) than for the polar 
and small hydrophobic amino acids (for which water should be better solvent).
Thus our model is seen to capture the substantial differences between 
the residue types in the correct manner.

\section{Conclusions}

We have constructed an $\alpha$-C based  molecular dynamics
model of IDP which, despite its simplicity, yields structural results
that are in a reasonable agreement with all-atom and experimental
data and allows for an access to significantly longer
timescales than in the all-atom simulations.\\

There is a category of IDPs that acquire structure on
binding with a substrate \cite{Kihara}. Our model is likely to fail
when using it to predict the specific structure that arises on binding.
The model requires a more stringent fine tuning for this purpose.
We have also found that the potentials used in our model cannot keep 
any structured  proteins in their native conformation, as equilibrium 
inter-residue distances are different. Thus applying our model to study, 
say, folding of the structured proteins leads to a fairly chaotic behavior.
In this case, using the structure-based models
is much more effective.\\

An alternative way to introduce specificity into the ss contacts
is to use the sums of statistically determined radii associated
with the residues \cite{Chwastyk} %as in ref. \cite{Chwastyk} 
instead of the pair-wise characteristics summarized in Table \ref{rmin}.
This approach would introduce merely 20 residue-related parameters.
It remains to be explored whether it provides an improvement over
the methodology presented here.\\

%The agreement presented here for monomers is a sufficient justification to 
%use presented approach in studies of much larger biomolecular complexes, such as
%polyQ aggregates\cite{prime,Esposito} and gluten\cite{gluten}, where much less experimental data is available for comparison of the molecular details. \\

{\bf Acknowledgements}
We appreciate fruitful discussions with B. R\'o\.zycki as well as 
receiving help of G. Matyszczak and K. Wo{\l}ek with the calculations.
MC has received funding from
the National Science Centre (NCN), Poland, under grant No.~2014/15/B/ST3/01905
and by the EU Joint Programme in Neurodegenerative Diseases project
(JPND CD FP-688-059) through the NCN grant 2014/15/Z/NZ1/00037.
He was also supported by the NCN grant No.~2016/21/B/NZ1/00006.
The computer resources were supported by the PL-GRID infrastructure
and also financed by the European Regional Development Fund under the
Operational Programme Innovative Economy NanoFun POIG.02.02.00-00-025/09.

\clearpage

\begin{table}[h!]
\begin{center}
\begin{tabular}{|l|llllllllll|}
\hline
name & {\bf Gly} & {\bf Pro} & {\bf Gln} & {\bf Cys} & {\bf Ala} & {\bf Ser} & {\bf Val} & {\bf Thr} & {\bf Ile} & {\bf Leu}\\
type & - & - & P & P & H & P & H & P & H & H\\
$s$   & 0 & 0 & 2 & 3 & 3 & 2 & 4 & 2 & 5 & 5\\
$n_H$ & 0 & 0 & 0 & 2 & 1 & 0 & 4 & 0 & 4 & 4\\
$n_P$ & 0 & 0 & 2 & 2 & 1 & 2 & 1 & 2 & 2 & 2\\ \hline
$n_{max}$ & 23 & 27 & 79 & 11 & 24 & 58 & 9 & 24 & 6 & 13\\
ID$_{UniProt}$ & P10275 & Q9NP73 & Q156A1 & Q03751 & Q8R089 & O15417 & P32867 & Q869S5 & Q95US5 & Q80YA8\\
\hline
$\nu$ & 0.56 & 0.93 & 0.61 & 0.48 & 0.57 & 0.54 & 0.46 & 0.56 & 0.51 & 0.49 \\ \hline
&  &  &  &  &  &  &  &  &  & \\
\hline %\hline
name & {\bf Asn} & {\bf Asp} & {\bf Lys} & {\bf Glu} & {\bf Met} & {\bf His} & {\bf Phe} & {\bf Arg} & {\bf Tyr} & {\bf Trp}\\
type & P & P- & P+ & P- & H & P & H & P+ & H & H\\
$s$   & 2 & 2 & 2 & 2 & 4 & 2 & 6 & 2 & 4 & 5\\
$n_H$ & 0 & 0 & 0 & 0 & 1 & 0 & 4 & 0 & 2 & 4\\
$n_P$ & 2 & 2 & 2 & 2 & 1 & 2 & 2 & 2 & 2 & 3\\ \hline
$n_{max}$ & 46 & 45 & 11 & 33 & 7 & 23 & 9 & 14 & 6 & 6\\
ID$_{UniProt}$ & Q54XG7 & Q08438 & Q8CI03 & Q6PCN3 & Q01668 & Q6ZQ93 & Q3S2U2 & P38835 & Q9C5D3 & P86690\\ \hline
$\nu$ & 0.54 & 0.81 & 0.81 & 0.81 & 0.60 & 0.62 & 0.52 & 0.81 & 0.51 & 0.44 \\ \hline %\hline
\end{tabular}
\caption{Attributes of the residues. The first line (after the name of the residue)
indicates the amino acid class.
The subscripts + or -- indicate whether the amino acids are charged.
The second line shows the number ($s$) of the sidechain
contacts that the amino acid can make with all other residues.
The third and fourth lines lines show the  numbers of sidechain 
contacts that the amino acids can make with the hydrophobic ($n_H$)
and polar ($n_P$) residues. The fifth line list the sequential extensions
of the longest homopolymeric tracts made with the residue as found in the
reviewed part of UniProt database (SProt), with the highest level of certainty (PE=1, 
see \cite{UniProt}). The sixth line gives the corresponding accession code.
The seventh line lists the exponent ($\nu$) for the sequence-length dependence
($n$) of $R_g$. The lines further down are for the remaining 10 residues.
His is considered to be uncharged.}
\label{aaproperties}
\end{center}
\end{table}

\begin{table}
\begin{center}
\begin{tabular}{|l|l|l|l|l|l|l|l|l|l|l|l|l|l|l|l|l|l|l|}
\hline
 & {\bf Gln} & {\bf Cys} & {\bf Ala} & {\bf Ser} & {\bf Val} & {\bf Thr} & {\bf Ile} & {\bf Leu} & {\bf Asn} & {\bf Asp} & {\bf Lys} & {\bf Glu} & {\bf Met} & {\bf His} & {\bf Phe} & {\bf Arg} & {\bf Tyr} & {\bf Trp}\\ \hline
{\bf Gln} & 8.63 & & & & & & & & & & & & & & & & &\\ \hline
{\bf Cys} & 7.72 & 7.56 & & & & & & & & & & & & & & & &\\ \hline
{\bf Ala} & 7.39 & 6.97 & 6.42 & & & & & & & & & & & & & & &\\ \hline
{\bf Ser} & 7.64 & 6.97 & 6.53 & 6.65 & & & & & & & & & & & & & &\\ \hline
{\bf Val} & 7.81 & 7.56 & 7.06 & 7.17 & 7.65 & & & & & & & & & & & & &\\ \hline
{\bf Thr} & 7.77 & 7.40 & 6.94 & 6.97 & 7.54 & 7.30 & & & & & & & & & & & &\\ \hline
{\bf Ile} & 8.24 & 7.95 & 7.45 & 7.52 & 8.06 & 7.93 & 8.53 & & & & & & & & & & &\\ \hline
{\bf Leu} & 8.44 & 8.07 & 7.65 & 7.68 & 8.29 & 8.12 & 8.77 & 8.93 & & & & & & & & & &\\ \hline
{\bf Asn} & 8.19 & 7.49 & 7.02 & 7.18 & 7.54 & 7.46 & 7.96 & 8.14 & 7.74 & & & & & & & & &\\ \hline
{\bf Asp} & 8.15 & 7.18 & 6.73 & 6.99 & 7.22 & 7.19 & 7.65 & 7.86 & 7.50 & & & & & & & & &\\ \hline
{\bf Lys} & 8.69 & 7.83 & 7.26 & 7.73 & 7.69 & 7.79 & 8.16 & 8.39 & 8.11 & \underline{8.59} & & & & & & & &\\ \hline
{\bf Glu} & 8.41 & 7.45 & 7.04 & 7.41 & 7.50 & 7.51 & 7.97 & 8.20 & 8.00 & & \underline{8.90} & & & & & & &\\ \hline
{\bf Met} & 8.84 & 8.29 & 7.91 & 7.94 & 8.48 & 8.33 & 8.95 & 9.14 & 8.49 & 8.15 & 8.80 & 8.61 & 9.29 & & & & &\\ \hline
{\bf His} & 8.64 & 8.17 & 7.50 & 7.88 & 7.92 & 7.98 & 8.37 & 8.57 & 8.36 & 8.50 & 8.58 & 8.84 & 8.93 & 8.83 & & & &\\ \hline
{\bf Phe} & 8.95 & 8.50 & 8.17 & 8.24 & 8.69 & 8.58 & 9.11 & 9.34 & 8.65 & 8.51 & 8.79 & 8.75 & 9.55 & 8.98 & 9.73 & & &\\ \hline
{\bf Arg} & 9.26 & 8.24 & 7.99 & 8.27 & 8.31 & 8.50 & 8.76 & 8.98 & 8.87 & \underline{9.12} & & \underline{9.52} & 9.27 & 9.23 & 9.26 & & &\\ \hline
{\bf Tyr} & 9.27 & 8.26 & 8.02 & 8.36 & 8.39 & 8.58 & 8.78 & 9.02 & 8.96 & 9.35 & 9.04 & 9.48 & 9.28 & 9.38 & 9.56 & 9.51 & 9.34 &\\ \hline
{\bf Trp} & 9.58 & 8.95 & 8.65 & 8.75 & 9.22 & 9.14 & 9.57 & 9.79 & 9.11 & 9.10 & 9.21 & 9.48 & 10.02 & 9.66 & 10.17 & 9.82 & 10.08 & 10.85 \\
\hline
\end{tabular}
\caption{Average distances, in {\AA}, for the ss contacts as derived from the CATH database.
The underlined entries indicate 
a possibility of forming ionic bridges through the electrostatic attraction.}
\label{rmin}
\end{center}
\end{table}

\clearpage

\end{document}